\documentclass[10pt, letterpaper]{IEEEtran}
\IEEEoverridecommandlockouts
\usepackage{cite}
\usepackage{amsmath,amssymb,amsfonts}
\usepackage{algorithmic}
\usepackage{graphicx}
\usepackage{textcomp}
\usepackage{xcolor}
\usepackage{calrsfs}
\usepackage{mathtools}
\usepackage{listings}
\DeclareMathAlphabet{\pazocal}{OMS}{zplm}{m}{n}
\def\BibTeX{{\rm B\kern-.05em{\sc i\kern-.025em b}\kern-.08em
    T\kern-.1667em\lower.7ex\hbox{E}\kern-.125emX}}
\usepackage{tikz}
\usepackage{multirow}
\usetikzlibrary{pgfplots.groupplots}
\usepackage{pgfplotstable}
\usepackage{pgfplots}
\usepackage{pgf}
\usepackage{datatool}
\usetikzlibrary{positioning}
\usepackage{xfp}
\usepackage{subcaption}
\AtBeginEnvironment{tabular}{\footnotesize}
\usepackage[inline, shortlabels]{enumitem}
\setlist{itemjoin ={,\enspace},itemjoin* = { and\enspace}}
\definecolor{vgreen}{RGB}{104,180,104}
\definecolor{vblue}{RGB}{49,49,255}
\definecolor{vorange}{RGB}{255,143,102}

\lstdefinestyle{verilog-style}
{
	language=Verilog,
	basicstyle=\scriptsize,
	keywordstyle=\color{vblue},
	identifierstyle=\color{black},
	commentstyle=\color{vgreen},
	numberstyle=\tiny\color{black},
	numbersep=10pt,
	tabsize=8,
	moredelim=*[s][\colorIndex]{[}{]},
	literate=*{:}{:}1
}

\makeatletter
\newcommand*\@lbracket{[}
\newcommand*\@rbracket{]}
\newcommand*\@colon{:}
\newcommand*\colorIndex{%
	\edef\@temp{\the\lst@token}%
	\ifx\@temp\@lbracket \color{black}%
	\else\ifx\@temp\@rbracket \color{black}%
	\else\ifx\@temp\@colon \color{black}%
	\else \color{vorange}%
	\fi\fi\fi
}
\makeatother

\usepackage{float}
\usepackage{caption} 
\DTLsetseparator{ }

\usepackage{background}
\usepackage[usestackEOL]{stackengine}
\setstackgap{L}{\normalbaselineskip}
\SetBgContents{\color{blue}{\tiny \Longstack{PREPRINT - accepted at  IEEE International Conference on Computer Design (ICCD), 2021.}}}
\SetBgPosition{4.5cm,1cm}
\SetBgOpacity{1.0}
\SetBgAngle{0}
\SetBgScale{1.8}
\begin{document}
	\setlength{\abovedisplayskip}{3pt}
	\setlength{\belowdisplayskip}{3pt}
	
\bstctlcite{IEEEexample:BSTcontrol} 
\title{QFlow: \underline{Q}uantitative Information \underline{Flow} for Security-Aware Hardware Design in Verilog\vspace{-0.2cm}}

\author{\IEEEauthorblockN{Lennart M. Reimann, Luca Hanel, Dominik Sisejkovic, Farhad Merchant and Rainer Leupers}

\IEEEauthorblockA{\textit{Institute for Communication Technologies and Embedded Systems,} \textit{RWTH Aachen University, Germany}\\\{lennart.reimann, hanel, sisejkovic, merchantf, leupers\}@ice.rwth-aachen.de} \vspace{-1.18cm}}

\maketitle

\begin{abstract}

The enormous amount of code required to design modern hardware implementations often leads to critical vulnerabilities being overlooked. Especially vulnerabilities that compromise the confidentiality of sensitive data, such as cryptographic keys, have a major impact on the trustworthiness of an entire system. Information flow analysis can elaborate whether information from sensitive signals flows towards outputs or untrusted components of the system. But most of these analytical strategies rely on the non-interference property, stating that the untrusted targets must not be influenced by the source’s data, which is shown to be too inflexible for many applications. To address this issue, there are approaches to quantify the information flow between components such that insignificant leakage can be neglected. Due to the high computational complexity of this quantification, approximations are needed, which introduce mispredictions. To tackle those limitations, we reformulate the approximations. 
Further, we propose a tool QFlow with a higher detection rate than previous tools. It can be used by non-experienced users to identify data leakages in hardware designs, thus facilitating a security-aware design process.

\end{abstract}

\vspace{-0.4cm}
\section{Introduction}
\label{ch:introduction}
Many security issues are either caused accidentally by inexperienced hardware designers or by malicious modifications, such as hardware Trojans\cite{bhunia}\cite{sisejkovic}. In the rest of this work, those issues are referred to as vulnerabilities.
The vulnerabilities should be identified and removed at an early design stage, as later modifications result in higher costs or a longer time-to-market. They can be identified using suitable test models, but the required test models become more computationally extensive with the increasing design complexity\cite{atpg}, thus not all test cases can be elaborated in a reasonable time.
Additionally, theorem provers \cite{theorem_provers}, property checkers, and formal approaches give more certainty, but often require special expertise to be used properly and suffer from scalability issues. 

One of the critical features of hardware security is the confidentiality of data. Vulnerabilities endangering the confidentiality of sensitive signals, such as cryptographic keys, can be implemented easily and stay undetected in common test cases, as the trigger might not be present in the program code and data. Therefore, we focus on protecting the confidentiality in this work.
Information flow analysis is an evolving research area and used as a method to detect hardware vulnerabilities concerning the confidentiality of data carried by digital circuits. A variety of solutions exist working on virtual prototypes \cite{vp-ift} or Register-Transfer Level (RTL)\cite{sec-verilog, sapper, caisson} to analyze the flow of information of marked signals, either dynamically (tracking) or statically (analysis). As dynamic elaborations analyze the information flow at runtime, they can only guarantee the security of a hardware design for given test cases\cite{vp-ift}. Most of these tools rely on analyzing the flow of information for the non-interference property. But this property, which forbids the communication between trusted and untrusted hardware components, can often not be implemented, as many applications rely on such a communication.
State-of-the-art methods, such as QIF-Verilog \cite{qif-verilog}, use Quantitative Information Flow (QIF) analysis, which allows a quantification of the actual leakage using developed metrics that have been shown to be suitable to detect leakages in digital systems\cite{qif-foundations}\cite{qif-gleakage}. This analysis allows a new classification of leakage paths supporting the identification of data leakages. However, in this work, QIF-Verilog is shown to be unreliable in the identification of certain design vulnerabilities. Therefore, we introduce a more suitable methodology, QFlow, capable of detecting even the new vulnerabilities. 

The major contributions of this paper are:
\begin{enumerate*}[i) ]
	\item An operational tool that uses a suitable QIF metric, called QModel, to quantify the trustworthiness of a Verilog hardware description regarding the protection of its sensitive signals, e.g., cryptographic keys or user data
	\item A secure approximation of state-of-the-art QIF formulas that removes the possibility of false negative predictions, while only enabling false positives, as shown empirically
	\item User-adjustable input probabilities for the information theory equations on bit level.
\end{enumerate*}
\vspace{-0.2cm}
\section{Preliminaries}
\label{ch:preliminaries}
\vspace{-0.2cm}
\subsection{Attack Model}
We focus on vulnerabilities that are implemented during the RTL-design process. Those vulnerabilities can be exploited by an adversary after production. In this work, we assume that the attacker can observe the outputs and the non-secret inputs of the selected hardware modules at a random moment in time. These outputs might leak signals carrying sensitive information, such as user data, via leakage paths. Whether those observations are obtained by physical access to the design or via other untrusted hardware in the System-on-Chip (SoC) is not considered. Additionally, the intruder cannot set any of the input values of the module under attack. During the attack, the complete design structure is known to the adversary.
\vspace{-0.7cm}
\subsection{Information Flow Analysis}
The designer is interested in protecting certain parts of the hardware carrying sensitive data. For this purpose, previous works used labels to classify the sensitivity of hardware models and the data that is carried in them\cite{security_labels}.
In most information flow models the system of interest is separated into two partitions: High (H) and Low (L). The H label is commonly used to describe trusted hardware components processing sensitive data. 
When labeling the hardware components, the trusted components would be labeled H, while the remaining components are labeled L. These are abstract labels, but the labels are commonly applied by marking the component's hardware description with their respective label\cite{sec-verilog}. The labels H and L are propagated throughout the system related to the dependency of the signals. 
\vspace{-0.3cm}
\subsection{Quantitative Information Flow}
In many previous works \cite{refining_metrics} \cite{poster_qif_metric}, the vulnerability, a metric for the weakness or simplicity to guess the secret, has shown to be a reliable function when quantifying the information flow. 
The Bayes Vulnerability represents a special case of the g-vulnerability, when the adversary has only a single guess of the secret after one observation of the outputs and is only rewarded for a correct guess\cite{alvim_science_2020}\cite{qif-foundations}. For a secret $H$ with a probability distribution of $\pi_H$, the vulnerability is
\begin{equation}
V_1[\pi_H]=\max_{h\in\pazocal{H}} \pi_h,
\label{eq:prior_bayes_vul}
\end{equation}
where $\pi_h$ is the probability of the symbol $h\in H$.
In this work, we are interested in the leakage caused by an information flow from a secret $H$ to the outputs of a system $O$. Therefore, the Posterior Bayes Vulnerability $V_1[\pi_H \triangleright C]$ (PBV) has to be considered as well. The posterior vulnerability shows the vulnerability after observing the outputs $O$, if the secret bits $H$ have been applied to the deterministic channel $C$ ($\pi_H \triangleright C$):
\begin{equation}
V_1[\pi_H \triangleright C] = \sum_{o\in\pazocal{O}} \max_{h\in\pazocal{H}} J_{o,h}.
\label{eq:bayes_post_vuln}
\end{equation}
Here, $J$ represents the joint distribution for the variables in the index. The channel is an abstract definition of the hardware system.
When combining the two Bayes Vulnerabilities, the so-called Multiplicative Bayes Leakage $\pazocal{L}_1^\times$ is computed as:
\begin{equation}
	\pazocal{L}_1^\times := \frac{V_1[\pi \triangleright C]}{V_1[\pi]} .
	\label{eq:mul_bayes_leakage}
\end{equation}

As the leakage computation for a complete hardware description would be infeasible, approximations are needed.
\vspace{-0.2cm}
\section{QFlow}
\label{ch:methodology}
\vspace{-0.2cm}
\subsection{Basic Functionality}
Our QIF-metrics are used to quantify the information flow from the 'High' sources, the marked signals, via all the operations that are applied on the secret, to the 'Low' targets. All output ports of the top design are automatically marked as 'Low'. In contrast to QIF-Verilog, we preprocess the abstract syntax tree to allow a bit-wise analysis to track the information flow in more detail. We compute the leakages in several steps. First we build channels consisting of Boolean equations representing the partitioned hardware. Next, the input probabilities and PBV for the channels are computed and combined with the input leakages of the channel to compute the total leakages \textit{for every single secret bit}. 
\vspace{-0.3cm}
\subsection{Approximation Assumptions}
\label{ch:assumptions}
Certain assumptions are needed to reduce the amount of computations needed to determine leakage values for the secrets:
\begin{enumerate*}[i)]
	\item The secret and known bits are independent of each other
	\item Compute Probabilities: The inputs of every channel are independent of each other
	\item Compute Posterior Vulnerability: The inputs of every channel are independent of each other. Additionally for COMPARISON, ADDITION, and SUBTRACTION operations, the high inputs are assumed to be uniformly distributed
	\item Compute Leakage: The dependency of input leakages is appraised with Eq. \eqref{eq:input_leakages}.
\end{enumerate*}
As the low inputs can be observed by the adversary as well, they need to be integrated into the equation for the PBV. 
\vspace{-0.4cm}
\subsection{QModel}
The assumptions and appraisals lead to our mathematical model, called QModel, implemented in our tool QFlow. This model is further explained below.
As mentioned before, we modified the equation for the PBV (Eq. \eqref{eq:bayes_post_vuln}) to include the low input signals $L$. They are added to the equation as additional observable parameters---similar to the outputs $O$:
\begin{equation}
V_1[\pi \triangleright C|L]:=V_1[H|L,O] := \sum_{\substack{o\in\pazocal{O}\\ l \in \pazocal{L}}} \max_{h\in\pazocal{H}} J_{o,l,h},
\label{eq:our_posterior_vulnerability}
\end{equation}
with the joint probability distribution $J_{o,l,h}$,
\begin{equation}
\begin{aligned}
J_{o,h,l} &= \pi_{o|h,l}\cdot\pi_{h,l}.
\label{eq:our_joint_probability}
\end{aligned}
\end{equation}
A feasible way to compute the leakage in a channel cascade is needed. Thus, we derived the following statement for the leakage $\pazocal{L}_{X\rightarrow Z}$ (in bit) of a Markov chain:
\begin{equation}
\pazocal{L}_{X\rightarrow Z} := \pazocal{L}_{X\rightarrow Y} \cdot \frac{\pazocal{L}_{Y\rightarrow Z}}{\pazocal{L}_{Y\rightarrow Z,max}}.
\end{equation}
For a Markov chain of channels $X\rightarrow Y \rightarrow Z$, the input values $X$ are converted to the output $Z$ with $Y$ as an intermediate value. The data processing inequality states that no processing of Y can increase the information that Y has about X\cite{alvim_science_2020}. Thus, we weigh the leakage of channel $X\rightarrow Y$ with the leakage of the second channel, scaled with its maximum possible leakage. When inserting the multiplicative Bayes Leakage (Eq. \eqref{eq:mul_bayes_leakage}) and the maximum leakage for the second channel, the reciprocal of the Prior Bayes Vulnerability (Eq.~\eqref{eq:prior_bayes_vul}), we derive the following equation:

\begin{equation}
\begin{aligned}
\pazocal{L}_{X\rightarrow Z} &= \pazocal{L}_{X\rightarrow Y}\cdot\frac{\frac{V_1(\pi_Y\rhd C_{Y\rightarrow Z})}{V_1(\pi_Y)}}{\frac{1}{V_1(\pi_Y)}}\\ &= \pazocal{L}_{X\rightarrow Y}\cdot V_1(\pi_Y\rhd C_{Y\rightarrow Z}).
\end{aligned}
\end{equation}
The input leakage into a channel (now in bit) can be further approximated as stated before in the Section \ref{ch:assumptions} using \cite{boreale}
\begin{equation}
\pazocal{L} \le \sum_{i\in\text{Inputs}}\pazocal{L}_i,
\label{eq:input_leakages}
\end{equation}
resulting in our final equation for the leakage of a secret bit $H_j$ in a channel cascade $\pazocal{L}_{C,H_j}$. The probability distribution of the secret channel inputs is represented by $\pi_{HI}$ and is used to compute the overall PBV of that channel:
\begin{equation}
\pazocal{L}_{C,H_j} := \sum_{i\in\text{Inputs}}\pazocal{L}_{i,H_j} \cdot V_1(\pi_{HI}\rhd C).
\label{eq:channel_leakage}
\end{equation}
\vspace{-0.7cm}
\subsection{Toolflow}
\begin{figure}
	\centering
		\includegraphics[width=0.9\linewidth]{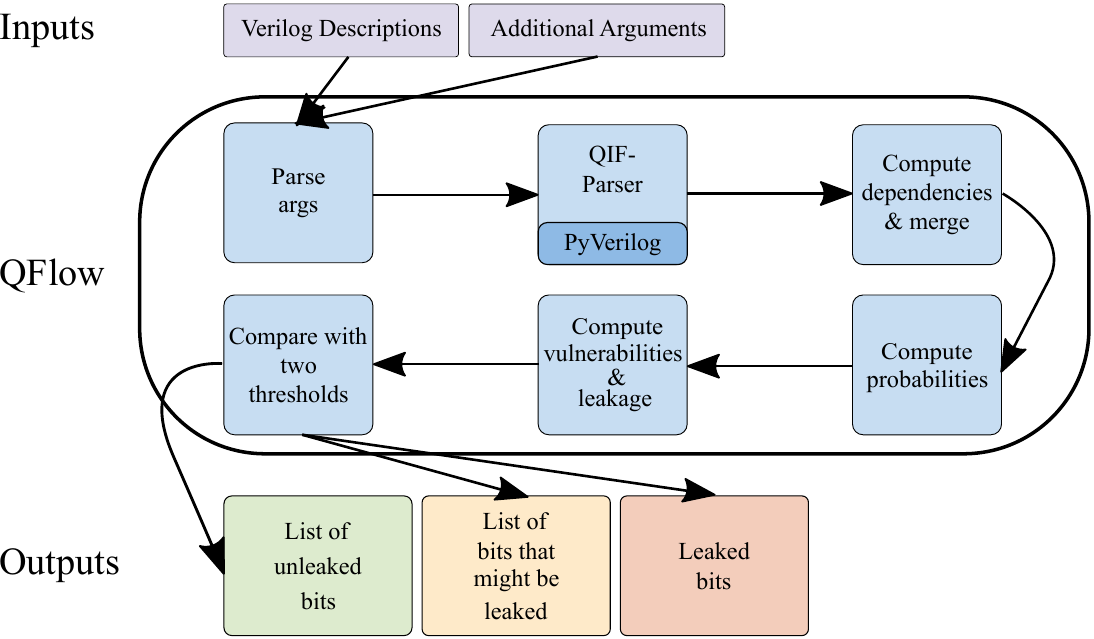}
	\caption{General toolflow of QFlow.}
	\label{fig:qif_toolflow}
	\vspace{-0.5cm}
\end{figure}
The general toolflow of QFlow is  illustrated in Fig. \ref{fig:qif_toolflow}. 
\subsubsection{ParseArgs}
Some parameters and the Verilog code are passed to the program. The signal that is supposed to stay secret is marked as 'High'.
\subsubsection{QIF-Parser}
The QIF-parser integrates the open-source tool PyVerilog, which is responsible for parsing Verilog and returning a graph of binds. A bind in such a graph mostly represents a single line of RTL code with terminal symbols, such as signals and constants as its leaves. 
As we intend to work on the bit level of the design, we need to process the bind tree. Then, each tree has an output bit as the root and only input signals and constants can be leaves.
An example for such a tree can be found in Fig. \ref{fig:tree_example}.
For the given circuit in Fig. \ref{fig:input_code}, the tree structure is illustrated for the first bit of the output in Fig. \ref{fig:computed_bind_tree}. The leaves (green) consist of the inputs of the circuit and constants. Operations (yellow) are connected to their outputs (orange) until an output, the root (red), is reached.
\subsubsection{Compute Dependencies \& Merge}
\begin{figure}
	\centering
	\hfill
	\begin{subfigure}[c]{0.46\linewidth}
		\vspace{-0.06cm}
		\begin{lstlisting}[style={verilog-style}, frame=single]
module example(
<@\textcolor{red}{High}@> input [1:0] i,
input low,
output [1:0] o);

wire k, s, t, u;

always @(*) begin
k = i[1] & low;
t = ~low;
s = i[0] & k;
u = i[0] & i[1];
o[0] = s ^ t;
o[1] = u | 0b1;
end
endmodule
		\end{lstlisting}
		\vspace{-0.2cm}
		\caption{Verilog code.}
		\label{fig:input_code}
	\end{subfigure}
	\hfill
	\begin{subfigure}[c]{0.49\linewidth}
		\includegraphics[width=\linewidth]{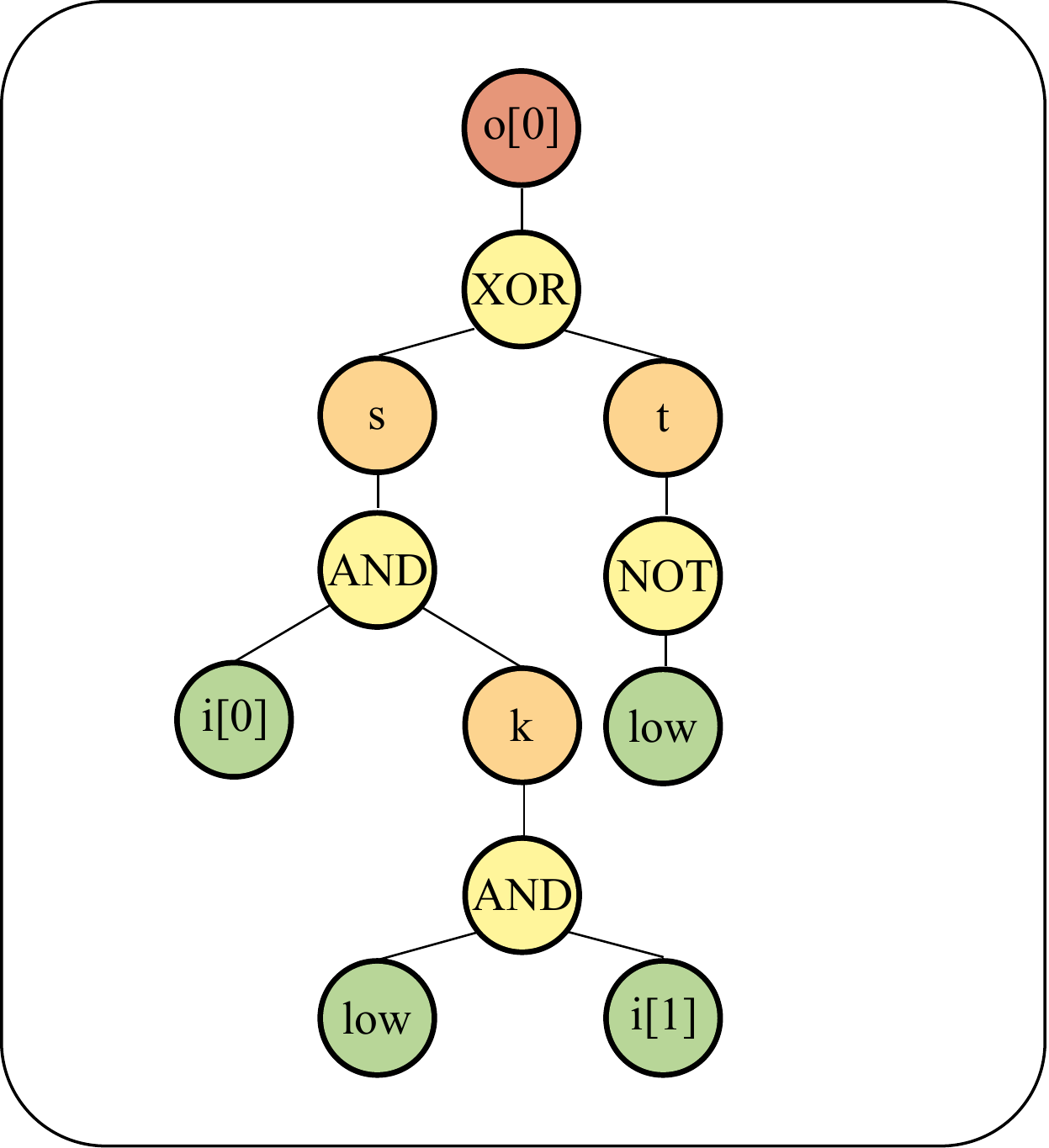}
		\caption{Parsed bind tree.}	
		\label{fig:computed_bind_tree}
	\end{subfigure}
	\begin{subfigure}{0.49\linewidth}
		\includegraphics[width=\linewidth]{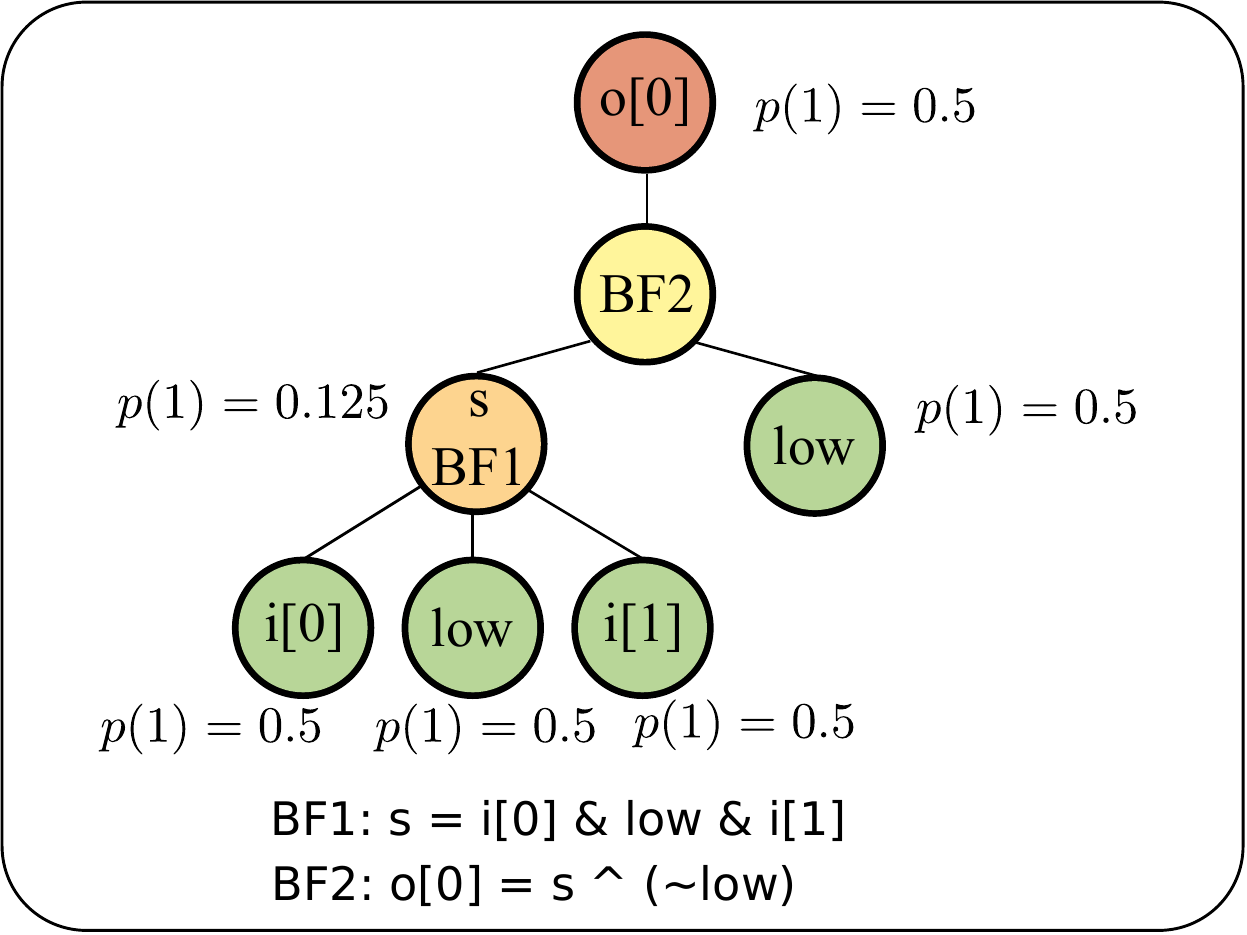}
		\caption{With probabilites.}
		\label{fig:computed_probabilities}
	\end{subfigure}
	\begin{subfigure}{0.49\linewidth}
		\includegraphics[width=\linewidth]{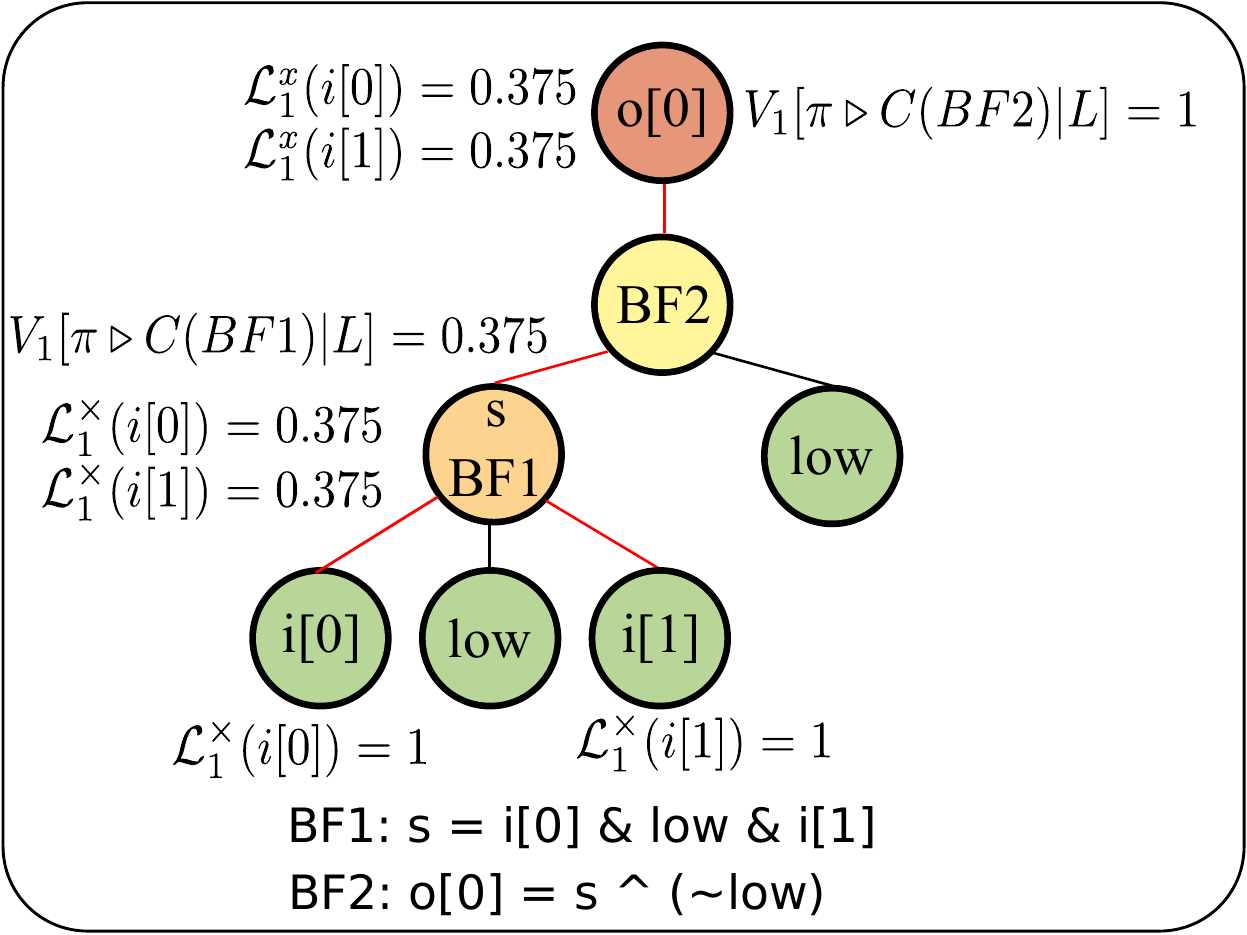}
		\caption{With vulnerab. \& leakages.}
		\label{fig:computed_leakages}
	\end{subfigure}
	\caption{Example for the computational steps for a small circuit.}
	\label{fig:tree_example}
	\vspace{-0.45cm}
\end{figure}

Afterwards, the tool computes the dependencies and starts merging nodes. The dependencies are needed to find loops in the tree structure to avoid merging them infinitely.
Merging is done to increase the channel sizes, thus reducing the number of channels that are cascaded. Every cascade of channels introduces an error in the computed leakage, due to the approximations done with the equations presented before. A negative error can be introduced due to the assumption of independent channel inputs, whereby the simple addition of input leakages introduces a positive error. In this paper, it is shown empirically that the positive error outweighs the negative one. The merging is stopped when the maximum number of input bits in the Boolean expression (\texttt{max\_channel\_inputs}) is reached. This is done to reduce the complexity for the next step, computing the probabilities. Furthermore, sequential branches are not merged as this may result in a security risk and misinterpretation of hardware vulnerabilities, which introduces an additional positive error. The example's tree was merged with \texttt{max\_channel\_inputs=3}, as shown in Fig. \ref{fig:computed_probabilities}. 
\subsubsection{Compute Probabilities}
The merged trees are forwarded to the 'Compute Probabilities' function. Here, the output probabilities of every channel input bit are computed. 
The probability of the outputs is calculated by multiplying the input probabilities for a given case because of the assumed independence. 
An example for the computation of the probabilities is shown in Fig. \ref{fig:computed_probabilities} with uniform input probabilities.
\subsubsection{Compute Vulnerability \& Leakage}
All the computed probabilities are written into the respective nodes. Next, it is possible to compute the posterior vulnerabilities for every channel using Eq. \eqref{eq:our_posterior_vulnerability}. All the example's computed probabilities are illustrated in Fig. \ref{fig:computed_probabilities}. For the computation of the PBV, the joint probability distributions need to be computed with Eq. \eqref{eq:our_joint_probability} for every channel.
Furthermore, the leakage of the input bits is set for every secret bit. If multiple inputs hold information about a secret bit, their leakage is added beforehand to appraise possible dependencies. From that point on, the leakage of every secret bit is multiplied with the PBV of the current channel. 
For every output bit, the leakages for all secret bits that influence it are computed by using the Eq. \eqref{eq:channel_leakage} over all channels starting at the leaves. Finally, the total leakage for every secret bit is computed by adding their leakage from every output bit. 
\subsubsection{Compare with two Thresholds}
Now that the leakage values for the different secret input bits are computed, we need to determine, which of the data paths are actually a vulnerability in the system. 
We determine a threshold and compare our computed value for every secret bit to it. 
\subsubsection{Classification}
After the accumulated leakages for every bit that is labeled with 'High' is computed and they have been compared to the thresholds, the leakage paths can be written out. 
This is done in three groups: Paths that leak (higher than detection threshold), paths that might leak (warning threshold), and paths that do not leak (below both thresholds). For the example, this is shown with the red connections in Fig. \ref{fig:computed_leakages}.
\vspace{-0.2cm}

\section{Evaluation}
\label{ch:experimentation}
For the evaluation, multiple open-source benchmarks were used to show the efficiency and flexibility of QFlow in detecting security vulnerabilities concerning the confidentiality of data. During the design process, the hardware designer is capable of elaborating the security of certain signals by marking them with 'High' inside his code. A second parameter that we alternate during the experimentation is the probability of the high input bits. This is done to show the influence of the probability on the leakage and illustrates the capability of the tool to emphasize the detection of hardware Trojans using controllable triggers. The same benchmarks are used to elaborate QIF-Verilog as well, thus they allow a comparison.
In the beginning, the required thresholds were determined by applying QFlow to two pipelined rounds of AES. This was done for uniformly distributed inputs, as they allow the highest leakage, leading to a more general threshold, while varying \texttt{max\_channel\_inputs}.
For most of the elaborations, AES and RSA benchmarks from Trust-Hub \cite{trusthub2} are used, which offer cryptographic accelerators in Verilog (and VHDL \cite{vhd2vl}).
They offer a variety of Trojans using either triggers or continuously write out the keys over additional output ports.
Furthermore, we implemented additional benchmarks to prove QFlow's functionality compared to QIF-Verilog.
\vspace{-0.3cm}
\subsection{Results}
At first, we analyzed a single AES benchmark for varying high input probabilities to illustrate the flexibility of QFlow and show the meaning of leakage. For the analysis, the probability of the 128-bit key was altered from 0 to 1 for all secret bits equally. The results (Fig. \ref{fig:mult_probs}) show that although the leakage for the system is 0 for the higher and lower probabilities, the attacker can still guess the keys as the Prior Bayes Vulnerability is at 1 bit. 
Even a solid hardware implementation cannot protect an insecure secret.

As mentioned before, the threshold was determined by applying QFlow to a Verilog implementation of two pipelined rounds of AES (by reducing one of the AES benchmarks). Two complete rounds of AES are supposed to lead to a full diffusion\cite{threshold}. Fig. \ref{fig:threshold} shows the minimum and average value of the 128-bit leakages for the keys with a varying \texttt{max\_channel\_inputs} value. As expected, the leakage drops, when more operations are merged into a single channel, reducing the error introduced by Eq. \eqref{eq:input_leakages} and the assumed independence of inputs. The smallest mean and min leakages that were computed were chosen as the threshold for the warning (min, $2.89154\cdot10^{-3}$) and threat (mean, $1.53939\cdot10^{-2}$) for the toolflow. The leakage did not drop after setting the maximum number of input values during the merge to 5. Thus, the value for \texttt{max\_channel\_inputs} is set to 5 for all following elaborations.
\begin{figure}
\begin{subfigure}{0.49\linewidth}
	\centering
	\begin{tikzpicture}[scale=0.5]
	\begin{axis}[
	xmin = 0, xmax=9,
	ymin = 0, ymax = 0.12,
	xlabel = \texttt{max\_channel\_inputs},
    ylabel style={yshift=0.4cm},
	ylabel = Leakage (bit),
	legend pos=north east,
	]
	\addplot file{2round_mean.txt};
	\addlegendentry{mean}
	\addplot file{2round_min.txt};
	\addlegendentry{min}
	\end{axis}
	\end{tikzpicture}
	\caption{Minimum and mean leakage.}
	\label{fig:threshold}
\end{subfigure}
\begin{subfigure}{0.49\linewidth}
	\centering
	\begin{tikzpicture}[scale=0.5]
	\begin{axis}[
	xmin = 0, xmax=1,
	ymin = 0, ymax = 40,
	xlabel = \texttt{probability of all secret bits being 1s},
	ylabel style={yshift=-0.4cm},
	ylabel = Leakage (bit),
	]
	\addplot file{mult_prob_aes_t1600.txt};
	\end{axis}
	\end{tikzpicture}
	\caption{Leakage over probabilities.}
	\label{fig:mult_probs}
\end{subfigure}
\caption{Leakage for two pipelined AES rounds varying \texttt{max\_channel\_inputs} (left). Leakage for AES-T100 AES varying probabilities \texttt{max\_channel\_inputs=5} (right).}
\vspace{-0.5cm}
\end{figure}
After the threshold was determined, the first analyses on the AES benchmarks were conducted. The results of these experiments can be seen in Fig. \ref{fig:t1200&1600}, illustrating the computed leakage for two different Trojans over the 128-bit key. The T-1200 benchmark leaks the first 8 bit of the key, which was XOR-ed with a value that can be determined by the intruder value. Thus, the XOR is not reducing the information content. Additionally, the Trojan is triggered continuously. This behavior is shown in Fig. \ref{fig:T1200}. The first 8 bits are leaked completely, while the other bit's leakages are low, due to the 10 rounds of AES. A second Trojan leaks the entire 128-bit key sequentially, when a certain condition is set. This condition and the sequential communication, reduces the likelihood of the attacker to gain information reducing the leakage to around 0.22 bits. For both benchmarks all leakages are detected and can be assigned to the actual secret bits. 
Table \ref{tab:AllTrojanResults} gives a summary of the experimental results for the AES and RSA benchmarks each containing a variety of Trojans. All leakages are detected, but one false positive detection and warning are given out for the RSA benchmarks. A short runtime of around 145 to 250s supports the claim that QFlow can assist designers in combination with their electronic design automation tools to allow a fast, yet security-aware design process. 
If secret bits are leaked entirely, it is detected as such with a leakage value of 1. 
If additional operations are executed on the values, or triggers reduce the probability of them being leaked, their leakage value also reduces, as they are less likely to be leaked, posing a smaller threat for the given attack scenario. 
\begin{table*}

	\centering
\caption{AES- and RSA-Trojan benchmark leakages \cite{trusthub1}\cite{trusthub2}("AL" = Average leakage).}
\vspace{-0.3cm}

\label{tab:AllTrojanResults}

	\begin{tabular}{c|c c||c c||c c||c||c}

		\hline
		\multirow{2}{*}{\textbf{Benchmark}} &   \textbf{\#Detected} & \textbf{Avg. Det.} & \textbf{\#FP Det.}& \textbf{\#FP Warn.} & \textbf{\#Unleaked} & \textbf{Avg. Sec.}  & \textbf{Time} & \textbf{Trojan Type}\\
		&   \textbf{/\#Actual} & \textbf{Leakage} & \textbf{/AL}& \textbf{/AL } & \textbf{/\# Actual} & \textbf{Leakage} &  (s) &\textbf{leaking information}\\\hline\hline
		AES-T100 & 8/8 & 1 & 0/- & 0/- & 120/120 & $2.66\cdot10^{-4}$ &  246 & Trigger:always; Payload:covert channel \\\hline
		AES-T200 & 8/8 & 1 & 0/- & 0/- & 120/120 & $2.66\cdot10^{-4}$ &  214 & Trigger:always; Payload:covert channel  \\\hline
		AES-T400 & 128/128 & 0.183 & 0/- & 0/- & 0/0 & - & 245 & Trigger:input; Payload:RF signal\\\hline
		AES-T700 & 8/8 & 1 & 0/- & 0/- & 120/120 & $2.66\cdot10^{-4}$ &  236 & Trigger:input; Payload:covert channel  \\\hline
		AES-T800 & 8/8 & 1 & 0/- & 0/- & 120/120 & $2.66\cdot10^{-4}$ &  232 & Trigger:input; Payload:covert channel  \\\hline
		AES-T900 & 8/8 & 1 & 0/- & 0/- & 120/120 & $2.66\cdot10^{-4}$ &  231 & Trigger:counter; Payload:covert channel \\\hline
		AES-T1000 & 8/8 & 1 & 0/- & 0/- & 120/120 & $2.66\cdot10^{-4}$ &  232 & Trigger:input; Payload:covert channel  \\\hline
		AES-T1100 & 8/8 & 1 & 0/- & 0/- & 120/120 & $2.66\cdot10^{-4}$ &  238 & Trigger:input; Payload:covert channel \\\hline
		AES-T1200 & 8/8 & 1 & 0/- & 0/- & 120/120 & $2.66\cdot10^{-4}$ &  233 & Trigger:counter; Payload:covert channel \\\hline
		AES-T1600 & 128/128 & 0.222 & 0/- & 0/- & 0/0 & - &  234 & Trigger:input; Payload:RF signal\\\hline
		AES-T1700 & 128/128 & 0.295 & 0/- & 0/- & 0/0 & - &  148 & Trigger:counter; Payload:RF signal\\\hline
		RSA-T100 & 33/32 & 0.5 & 1/0.023 & 1/0.006 & 30/32 & $6.3\cdot10^{-5}$ &  196 & Trigger:input; Payload:via Ciphertext \\\hline
		RSA-T300 & 33/32 & 0.5 & 0/0.023 & 1/0.023 & 30/32 & $6.3\cdot10^{-5}$ &  191 & Trigger:counter; Payload:via Ciphertext\\\hline
	\end{tabular}
\vspace{-0.3cm}
\end{table*}
\begin{figure}[t]
	\begin{subfigure}{\linewidth}
		\centering
		\begin{tikzpicture}[scale=0.8]
		\begin{axis}[ybar,ymode = log, log origin=infty, bar width = 0.01cm, axis x line=bottom, axis y line=left, ymin=0.0007, yscale = 0.2, ymax = 1, xtick={0,127}, yminorticks = false, axis line style={-}, xlabel={Secret bits}, ylabel={Leakage (bit)},ylabel style={xshift=-0.6cm},xlabel style={yshift=-0.6cm}]
		
		\addplot file{AES-T1200_m5_b2_reduced.txt};
		\addplot[red,sharp plot,update limits=true,line width=1pt] coordinates { (0,0.015393853187561035) (127,0.015393853187561035) };
		\addplot[orange,sharp plot,update limits=true,line width=1pt] coordinates { (0,0.00289154052734375) (127,0.00289154052734375) };
		\end{axis}
		\end{tikzpicture}
		\caption{Leakage of the AES-T1200 key bits.}
		\label{fig:T1200}
	\end{subfigure}
	\begin{subfigure}{\linewidth}
		\centering
			\begin{tikzpicture}[scale=0.8]
			\begin{axis}[ybar,ymode = log, log origin=infty, bar width = 0.01cm, axis x line=bottom, axis y line=left, ymin=0.0007, yscale = 0.2, ymax = 1, xtick={0,127}, yminorticks = false, axis line style={-}, xlabel={Secret bits}, ylabel={Leakage (bit)},ylabel style={xshift=-0.6cm},xlabel style={yshift=-0.6cm}]
			
			\addplot file{AES-T1600_m5_b2_reduced.txt};
			\addplot[red,sharp plot,update limits=true,line width=1pt] coordinates { (0,0.015393853187561035) (127,0.015393853187561035) };
			\addplot[orange,sharp plot,update limits=true,line width=1pt] coordinates { (0,0.00289154052734375) (127,0.00289154052734375) };
			\end{axis}
			\end{tikzpicture}
		\caption{Leakage of the AES-T1600 key bits.}
		\label{fig:t1600}
	\end{subfigure}
	
	\caption{Leakages (Thresholds: Det. (Red) and Warn. (Orange)).}
	\label{fig:t1200&1600}
	\vspace{-0.6cm}
\end{figure}
\vspace{-0.3cm}
\subsection{Additional Benchmarks}
As mentioned before, we implemented some benchmarks and used the designs of the AES-T series from Trust-Hub  as a basis. The modifications to the designs are explained in Fig. \ref{fig:NewBenchmarks}, staying with the naming convention and introducing T2100, T2200, and T2300. 
As QIF-Verilog ignores data dependencies and increases its uncertainty value depending on what operation is conducted on the secret, disregards which signals are combined, certain vulnerabilities are introduced. We rebuilt QIF-Verilog \cite{qif-verilog} and tested it on those benchmarks. The results can be seen in Fig. \ref{fig:NewBenchmarks} d). Trojans that use operations that deconcatenate the key with itself followed by a concatenation, are not detected, as the tool identifies the operation as an increase in the uncertainty caused by the deconcatenation, although the value is not changed at all. The other Trojans are not explained further but are described in detail in Fig. \ref{fig:NewBenchmarks}a-c. QFlow was able to detect all those Trojans using the adapted equations and the merging approach to identify dependencies and quantify the leakage more accurately. 
\begin{figure*}[t]
	\centering
	\hspace{0.1cm}
	\begin{subfigure}{0.18\textwidth}
		\begin{lstlisting}[style={verilog-style}, basicstyle=\tiny, frame=single]
//In TSC.v
module TSC(
input rst,
input clk,
input [127:0] key,
output [63:0] load
);
reg [63:0] load;
reg [63:0] tmp0,tmp1,
           tmp2,tmp3;
           
genvar i;
generate
for (i=0; i < 64; i=i+1)
begin
always @ (posedge clk) 
begin
 tmp0[i] <= key[i];
 tmp1[i] <= tmp0[i];
 tmp2[i] <= tmp1[i];
 tmp3[i] <= tmp2[i];
 load[i] <= tmp3[i];
end
end
endgenerate

endmodule

		\end{lstlisting}
\caption{AES-T2100}
	\end{subfigure}\hfill
\begin{subfigure}{0.18\textwidth}
	\begin{lstlisting}[style={verilog-style}, basicstyle=\tiny, frame=single]
//In TSC.v
module TSC(
input rst,
input clk,
input [127:0] key,
output [63:0] load
);

reg [63:0] load;
reg [63:0] tmp0,tmp1;
reg [63:0] tmp2,tmp3;
reg [63:0] tmp4,tmp5;


always @ (posedge clk)
begin
 tmp0 <= key[63:0] 
         & key[63:0];
 tmp1 <= key[63:0] 
         | key[63:0];
 tmp2 <= tmp0 ^ tmp1;
 tmp3 <= tmp0 | tmp1;
 tmp4 <= tmp2 ^ tmp3;
 tmp5 <= tmp3 & tmp3;
 load <= tmp4 | tmp5;
end
endmodule
	\end{lstlisting}
\caption{AES-T2200}
\end{subfigure}
\hfill
\begin{subfigure}{0.20\textwidth}
	\begin{lstlisting}[style={verilog-style}, basicstyle=\tiny, frame=single]
//In top.v
TSC Trojan (rst, clk, 
            key, state,
            Capacitance); 
//In TSC.v
module TSC(
input rst,
input clk,
input [127:0] key,
input [127:0] in,
output [63:0] load
);
reg [63:0] load;
reg [127:0] tmp0,tmp1;
reg [127:0] tmp2,tmp3;
reg [127:0] tmp4;

always @ (posedge clk)
begin
 tmp0 <= in ^ key;
 tmp1 <= tmp0 ^ in;
 tmp2 <= tmp1 ^ in;
 tmp3 <= tmp2 ^ in;
 tmp4 <= tmp3 ^ in;
 load <= tmp4[63:0];
end
endmodule
\end{lstlisting}
\caption{AES-T2300}
\end{subfigure}\hfill
\begin{subfigure}{0.35\textwidth}
\vspace{1.2cm}
\begin{tabular}{ c| c| c| c }
	
	\hline
	& \multicolumn{3}{c}{\textbf{AES-Benchmarks}} \\\hline
	& T2100 & T2200 & T2300 \\ \hline
	\multicolumn{4}{c}{\textbf{QIF-Verilog (rebuilt)}}\\ \hline
	 \textbf{Time} (s) & 24.2 & 23.7 & 22.1 \\\hline
     \textbf{Accu. Uncertainty}& 379 & 320 & 384.0 \\ \hline
     \textbf{Threshold} & 318.72 &318.72 & 318.7 \\ \hline
	 \textbf{Detec.?}& No &No & No \\ \hline
	 \multicolumn{4}{c}{\textbf{QFlow}}\\ \hline
	 \textbf{Time} (s)& 298 &297 & 292 \\ \hline
	 \textbf{Total Leakage} & 64.03 & 64.03 & 64.03  \\ \hline
	 \textbf{Detec.?}& Yes &Yes & Yes \\ \hline
\end{tabular}
\vspace{1.3cm}
\caption{Leakages of new benchmarks.}
\end{subfigure}
\caption{New benchmarks to show the vulnerabilities in QIF-Verilog. Three Trojans leaking 64 bit.}
\label{fig:NewBenchmarks}
\vspace{-0.5cm}
\end{figure*}
\textit{Limitations: }
As mentioned before, the attack model for the designed QIF-model states that the attacker cannot set any input values, but can only observe the output once at a random moment. Knowing that the designer is not aware or certain about the design's vulnerability, they cannot know the inputs or triggers that make the circuit most vulnerable. 
This reduces the computed leakage compared to the actual leakage for highly unlikely leakage paths. 
Furthermore, hardware Trojans or unintentional vulnerabilities that leak a low amount of information over a longer period of time might not be detected. 
\vspace{-0.5cm}
\section{Conclusion}
\label{ch:conclusion}
In this publication, we introduced QFlow, a tool allowing the hardware designer to create a more security-aware design process using Quantitative Information Flow. It was shown that by using a suitable approximation, the hardware design can be separated into several channels to reduce the computational complexity, when independence of the channel's inputs is assumed. However, this dependency needs to be considered for merging leakage paths. The tool was proven to be more reliable for the defined attack model than the state-of-the-art tools. A simple usage and the acceptable computation time, support the claim that QFlow allows a security-aware design process that can be conducted by inexperienced users.
\vspace{-0.3cm}

\bibliographystyle{IEEEtran}
\bibliography{IEEEabrv,bibtexentry}

\end{document}